\documentclass[aps,prb,reprint,superscriptaddress]{revtex4-1}
\usepackage{amsmath}
\usepackage{amsfonts}
\usepackage{graphicx}
\usepackage{color}
\newcommand{\gf}[3]
{\langle\langle  #1 ;\, #2 \rangle\rangle_{#3}}
\newcommand{\ud}{\mathrm{d}}
\newcommand{\ket}[1]{\left| #1 \right>}

\newcommand{\braket}[2]{\left< #1 \right|\left. #2 \right>}

\newcommand{\expect}[3]{\big< #1 \big| #2 \big| #3 \big>}
\newcommand{\xpect}[1]{\left<  #1 \right>}

\newcommand{\abs}[1]{\left| #1 \right|}
\begin{document}

\title{Capacitive interactions and Kondo effect tuning in double quantum impurity systems}
%Kondo effect tuning by capacitive interactions in a double quantum dot system}

\author{David A. Ruiz-Tijerina}
\affiliation{Department of Physics and Astronomy and Nanoscale and Quantum 
Phenomena Institute, Ohio University; Athens, Ohio 45701--2979, 
USA}\affiliation{Instituto de F\'isica, Universidade de S\~ao Paulo, C.P. 
66318, 05315-970 S\~ao Paulo, SP, Brazil}

\author{E. Vernek}
\affiliation{Instituto de F\'isica, Universidade Federal de Uberl\^andia, 
Uberl\^andia, MG 38400-902, Brazil}\affiliation{Instituto de F\'isica de S\~ao Carlos, Universidade de S\~ao Paulo,
S\~ao Carlos, S\~ao Paulo, 13560-970, Brazil}

\author{Sergio E. Ulloa} \affiliation{Department of Physics and Astronomy 
and Nanoscale and Quantum Phenomena Institute, Ohio University; Athens, 
Ohio 45701--2979, USA}\affiliation{Dahlem Center for Complex Quantum 
Systems, Freie Universit\"at Berlin, 14195 Berlin, Germany}

\date{\today}

\begin{abstract}
We present a study of the correlated transport regimes of a double 
quantum impurity system with mutual capacitive interactions.  Such system
can be implemented by a double quantum dot arrangement or by a quantum dot and nearby 
quantum point contact, with independently connected sets of metallic terminals. 
Many--body spin correlations 
arising within each dot--lead subsystem give rise to the Kondo effect under
appropriate conditions. The 
otherwise independent Kondo ground states may be modified by the capacitive 
coupling, decisively modifying the ground state of the double quantum impurity system.
We analyze this coupled system through variational 
methods and the numerical renormalization group technique. Our 
results reveal a strong dependence of the coupled system ground state on 
the electron--hole asymmetries of the individual subsystems, as well as on their 
hybridization strengths to the respective reservoirs. The  
electrostatic repulsion produced by the capacitive coupling produces an effective shift of the 
individual energy levels toward 
higher energies, with a stronger effect on the `shallower' subsystem (that closer to 
resonance with the Fermi level), potentially pushing it out of the Kondo 
regime and dramatically changing the transport properties of the system.
The effective remote gating that this entails is found to depend nonlinearly on the 
capacitive coupling strength, as well as on the independent subsystem levels. 
The analysis we present here of this mutual interaction should be important to fully characterize
transport through such coupled systems.
\end{abstract}

\maketitle

\section{Introduction}\label{sec:intro}
Dramatic advances in fabrication techniques and control of nanostructures 
have led to a deeper understanding of the behavior of solid--state systems 
at the nanoscale. The transition from simple structures to more complex 
arrangements of fundamental building blocks has allowed the study of 
systems that exhibit rich physical behavior involving charge 
and spin degrees of freedom in many--body states; many of these structures
also exhibit potential for technological applications.

Interesting examples of such complex architectures are arrays of quantum dots 
(QD)---nanostructures with discrete energy spectra that act effectively as 
zero--dimensional quantum objects, containing one or few electrons. 
Originally built on semiconductor heterostructures, QDs have been 
implemented in a variety of systems, including carbon nanotubes and 
semiconductor nanowires.\cite{bjork_nanolett_2004}  It is possible to control 
the state of the QDs in the complex by a variety of external probes--- 
for instance, source--drain bias voltages, gate voltages, 
magnetic fields and even mechanical deformations.\cite{parks_science_2010} 
Systems consisting of 
two,\cite{holleitner_science_2002,hubel_apl_2007,hubel_prl_2008} 
three\cite{gaudreau_apl_2009,gaudreau_natphys_2012} or more QDs have been 
built, which exhibit many interesting properties, such as coherent 
electron tunneling\cite{holleitner_science_2002,vanderwiel_rmp_2002, 
holleitner_prb_2004} and novel many--body ground states.

\begin{figure}[h]
\begin{center}
\includegraphics[scale=0.24]{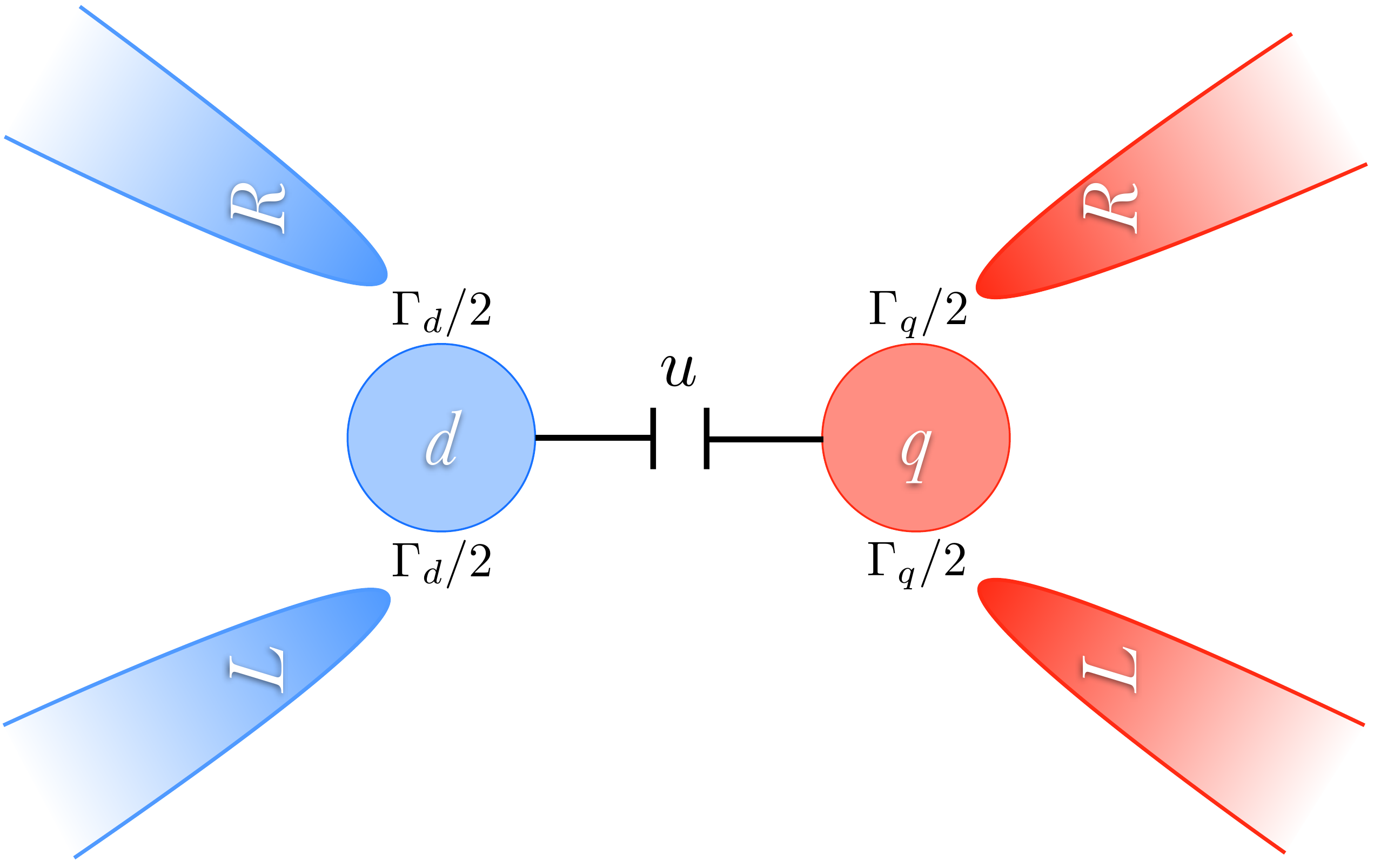}
 \caption{(Color online) Illustration of the model system studied: 
two quantum impurities, $d$ and $q$, capacitively coupled with interaction 
strength $u$. Each impurity $i(=d,\,q)$ is coupled to its own set of 
metallic terminals with hybridization energy $\Gamma_i/2$.}
\label{fig:double_dot_model}
\end{center}
\end{figure}

The Kondo 
effect\cite{kondo_1964,ng_prl_1988,glazman_jetplett_1988,Glazman2001} is a 
paradigmatic many--body phenomenon that has been repeatedly observed in QDs 
since the late 1990s,\cite{goldhaber_gordon_1998} and whose understanding is of 
fundamental importance in condensed matter physics.\cite{hewson_kpthf} It 
arises due to spin or 
pseudospin\cite{borda_prl_2003,hubel_prl_2008,okazaki_prb_2011} 
correlations between a quantum impurity and the itinerant electrons in the 
metallic reservoir in which it is embedded. The important role of the Kondo 
effect in the behavior of complex nanoscale systems has been demonstrated 
time and again: It has been observed in measurements of the dephasing by a 
quantum point contact (a charge detector on a nearby 
QD),\cite{avinun-kalish_prl_2004} while an unusual variety of Kondo 
effect with SU(4) symmetry was theoretically 
predicted\cite{galpin_DQD1,galpin_DQD2} and recently observed 
experimentally\cite{amasha_prl_2013,keller_nature_phys_2014} in a system of 
two identical, capacitively coupled QDs.  Many other examples exist.\cite{vanderwiel_rmp_2002} 
Capacitively coupled QDs have been recently predicted to exhibit a spatial 
rearrangement of the screening cloud under the influence of a magnetic 
field.\cite{vernek_arxiv_1308.4746}

In this work we present a study of the correlated transport regimes of a 
system of two independently contacted quantum impurities, each possibly in the Kondo regime,
and coupled to each other through capacitive interactions (see 
Fig.\ \ref{fig:double_dot_model}). We aim to understand how the Kondo correlations 
in one of them are affected by the presence and controlled variation of the other
subsystem. To 
this end, we carry out an analysis of the coupled system within the Kondo 
regime. The study utilizes a variational, as well as a numerical approach 
to the problem. The variational analysis gives us interesting insights into 
the effects of the capacitive coupling on the ground state of the coupled 
QD system. We further evaluate dynamic and thermodynamic properties of the 
system using the numerical renormalization group (NRG) method. We put 
especial emphasis on the conductances of both subsystems, which can be 
measured experimentally.

Our results show that the effects of the capacitive coupling can be 
absorbed into a positive shift of the local energies of the impurities, as 
one could anticipate. However, for the generic case of non--identical 
subsystems, the size of the effective shifts strongly depend on the 
relative magnitudes of the level depths with respect to the Fermi level, 
and the hybridization of each impurity to its electronic reservoirs. 
Moreover, the asymmetric shifts for effectively large capacitive coupling 
can drive the `shallower' subsystem into a mixed--valence regime, or even 
further away from the Kondo regime, resulting in interesting behavior of the 
relative Kondo temperature(s) and response functions, including the 
conductance levels through the separately--connected subsystems. 
Interestingly, a suppression of the unitary conductance produced by the 
Kondo effect is observed in the shallow impurity when the level shift 
produces a quantum phase transition out of the Kondo regime. These changes 
illustrate an interesting modulation of the Kondo correlations in one 
subsystem by a purely capacitive coupling, which could perhaps be useful in 
more general geometries.

In fact, all of the regimes presented in this paper are already accessible 
to experiments with double quantum dots, and may be of relevance for other 
types of quantum impurity systems as well. One can think in particular of 
the often used configuration, where a charge detector [a properly biased 
quantum point contact (QPC)] is placed in close proximity to an active QD. 
It is known that a QPC exhibits Kondo correlations,\cite{rejec_nature_2006} 
which will be clearly affected by the capacitive coupling to the dot.
With this in mind, in what follows we refer to the two independently connected impurities as 
the QD ($d$) and the QPC detector ($q$), and consider regimes where they can be seen as 
spin--$\frac{1}{2}$ quantum impurities coupled to their corresponding set of current 
leads. 

\section{Model}\label{sec:model}

We model the two subsystems as single--level Anderson Hamiltonians 
of the form
\begin{subequations}
\begin{equation}\label{eq:SIAMd}
\begin{split}
H_{d} =& \varepsilon_{d}\,n_d + U_d \, n_{d\uparrow}n_{d\downarrow} + 
\sum_{\mathbf{k}\sigma}\varepsilon_{k\sigma}n_{d\mathbf{k}\sigma}\\
&+ V_d\sum_{\mathbf{k}\sigma}\left(d_{\sigma}^{\dagger} 
c_{d\mathbf{k}\sigma } + \text{H. c.}\right),
\end{split}
\end{equation}
\begin{equation}\label{eq:SIAMq}
\begin{split}
H_{q} =& \varepsilon_{q}\,n_q + U_q \, n_{q\uparrow}n_{q\downarrow} + 
\sum_{\mathbf{k}\sigma}\varepsilon_{k\sigma}n_{q\mathbf{k}\sigma}\\
 &+V_q\sum_{\mathbf{k}\sigma}\left(q_{\sigma}^{\dagger}c_{q\mathbf{k} 
\sigma} + \text{H. c.}\right),
\end{split}
\end{equation}
\end{subequations}
 where the subindices $d$ and $q$ indicate the QD and the QPC, 
respectively, and $\varepsilon_d$ and $\varepsilon_q$ are the energies of 
the corresponding local levels; the number operators are given by 
$n_i=\sum_{\sigma}n_{i\sigma}$, with $\sigma=\uparrow,\,\downarrow$, and 
$U_i$ is the energy cost of double occupancy of level $i=d,\,q$ due to 
intra--impurity Coulomb interactions. The number operators 
$n_{i\mathbf{k}\sigma}$ give the occupation of the state of momentum 
$\mathbf{k}$ and spin projection $\sigma$ of the metallic terminal coupled 
to dot $i$. We consider here a band of half--bandwidth $D$ for each lead, 
with a flat density of states 
$\rho(\varepsilon)=\Theta(D-|\varepsilon|)/2D$, where $\Theta(x)$ is the 
Heaviside function. Assuming that the two terminals attached to each 
impurity are identical, we define the symmetric operators 
$c_{i\mathbf{k}\sigma}=\left( c_{iL\mathbf{k}\sigma} + 
c_{iR\mathbf{k}\sigma} \right)/\sqrt{2}$--- with indices $L$ and $R$ for 
the left and right terminals, respectively--- to which the impurity couples 
exclusively. The full Hamiltonian is thus
\begin{equation}\label{eq:fullH}
	H = H_{d} + H_{q} + u\,n_{d}n_{q},
\end{equation}
where the mutual capacitive coupling is parameterized by the energy $u>0$. 
The extra energy cost of simultaneous occupation of both impurities will 
produce a competition between their otherwise separate ground states: In 
the case of $\varepsilon_{i}<0$ and $\Gamma_i = \pi |V_i|^2/(2D) \ll 
\abs{\varepsilon_i}$, for both $i=d,\,q$, it would be favorable for each 
impurity to be singly occupied. The capacitive coupling, however, raises 
the energy of this coupled configuration and, depending on the magnitude of 
$u$, another occupancy may be more energetically favorable.

For interacting quantum impurities ($U_i > 0$) within this parameter 
regime, the Kondo effect\cite{kondo_1964,goldhaber_gordon_1998} takes place 
for temperatures below a characteristic temperature scale $T_K^i$. However, as 
we find below, the added energy cost of having a charge in a given impurity 
may drive the other impurity out of the Kondo regime at a critical coupling, as 
the mixed--valence regime is reached. 

In order to understand the competition between different impurities' 
ground states, we use the variational approach described in Appendix 
\ref{app:variational}. Because there is no charge exchange between the two 
subsystems, the effect of the capacitive coupling on each impurity can be 
absorbed into a level shift of the form $\varepsilon_{i} \rightarrow 
\varepsilon_{i} + \Lambda_{i}$, with each $\Lambda_i>0$ depending on the 
parameters of both quantum impurities, and on $u$. This shift can then be 
extracted variationally from a proposed wave function based on the form 
expected in the strong-coupling fixed point (SCFP) of the uncoupled 
systems,\cite{krishnamurthy_1980_1,krishnamurthy_1980_2} where the ground 
state has a many--body singlet structure. The level shifts $\Lambda_d$ and 
$\Lambda_q$ can be calculated by minimizing the energy of the coupled 
system. Within each subsystem, the corresponding level shift $\Lambda_i$ 
can be interpreted as an external gate voltage that raises the dot level, 
and thus determines the nature of its ground state.\cite{ferreira_prb_2011} 

In what follows, we describe the results of both the variational and NRG 
approaches, and explore the resulting physics and measurable consequences 
of the capacitive coupling. The NRG calculations fully capture the 
many--body correlations of the problem, and allow us to reliably calculate 
expectation values of operators, as well as the dynamic and thermodynamic 
properties of the system.

\begin{figure}[t]
\begin{center}
\includegraphics[scale=0.58]{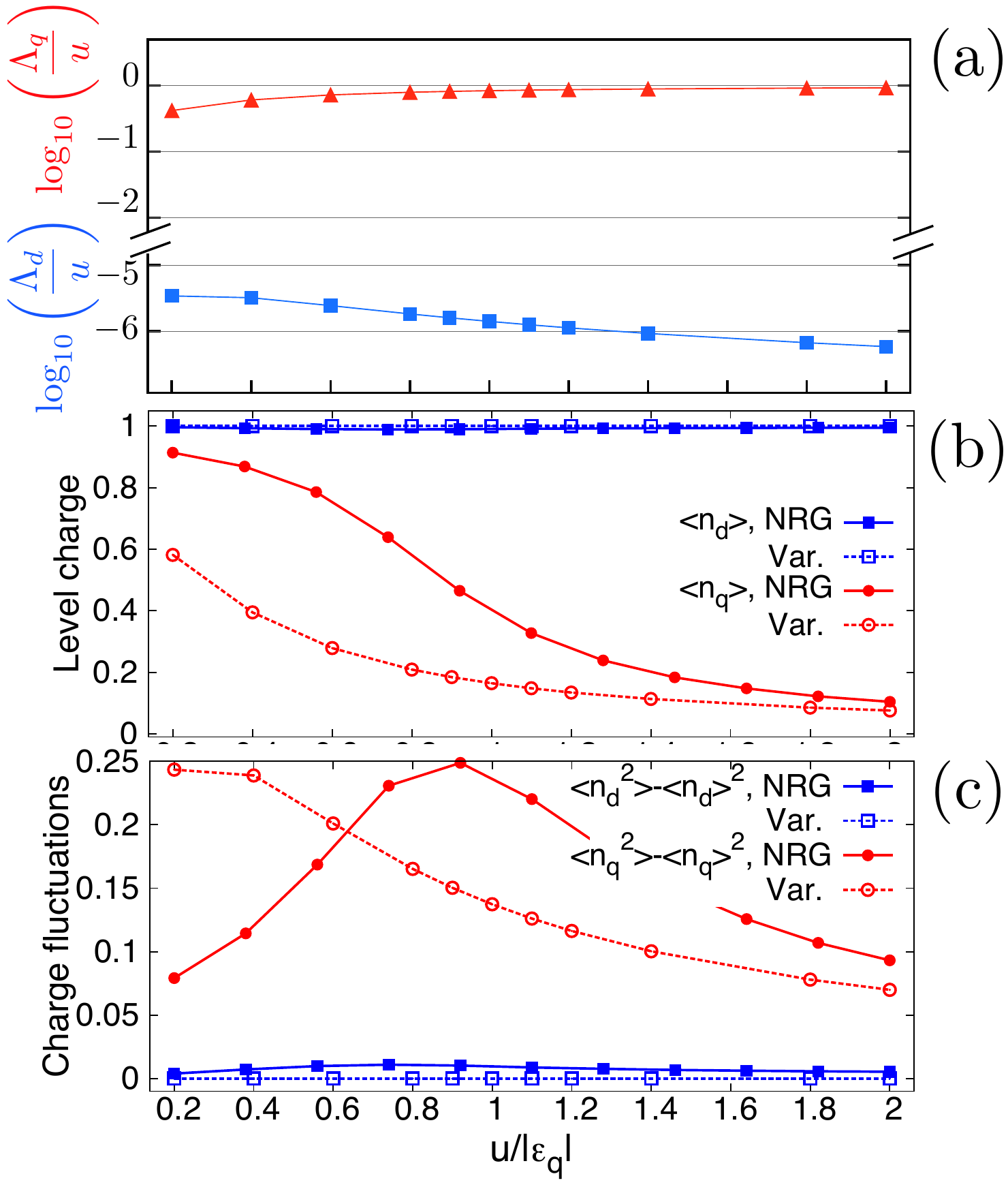}
 \caption{(Color online) (a) External gating of the impurity levels, 
$\Lambda_d$ and $\Lambda_q$ for identical couplings ($\Gamma_d = \Gamma_q = 
0.002\,D$), but asymmetric impurities, with $\varepsilon_d = -0.025\,D$, 
$\varepsilon_q = -0.010\,D$ and varying $u$, as obtained variationally. (b) 
Charge and (c) charge fluctuations of both impurities evaluated using the 
variational method and NRG, showing how impurity $q$ is depleted as the 
capacitive coupling $u$ is increased, while impurity $d$ remains 
essentially unchanged. The NRG calculations were carried out with local 
Coulomb interaction strengths of $U_d = U_q = 0.05\,D$.}
\label{fig:lambdas_vs_u}
\end{center}
\end{figure}

\begin{figure*}
\begin{center}
\includegraphics[scale=0.56]{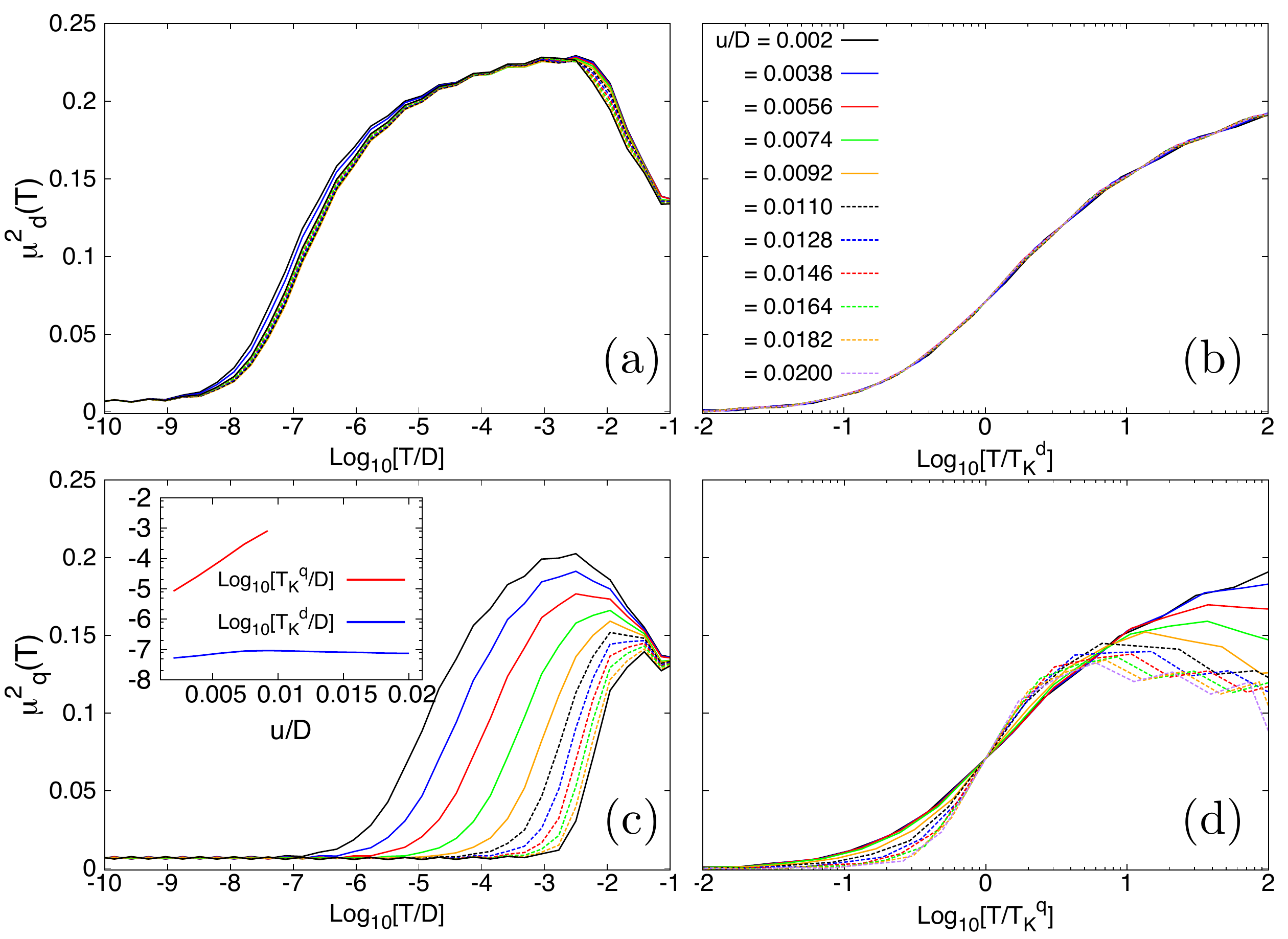}
 \caption{(Color online) Magnetic--moment--squared of impurities (a) $d$ 
and (c) $q$ as functions of the temperature, for different values of the 
capacitive coupling $u$, as shown in the legend in panel (b); all panels 
show results for the same set of $u$ values. (b) and (c) show the same quantities 
as functions of temperature, normalized to the Kondo temperature of each 
curve. The inset of (c) shows the dependence of the Kondo temperatures on 
$u$. All results obtained from NRG calculations. Parameters correspond to 
those of Fig.\ \ref{fig:lambdas_vs_u}. Notice that for $u \gtrsim 0.01\,D$ 
impurity $q$ is pushed away from the Kondo regime, first into 
mixed--valence, eventually reducing $n_q$ well below unity (see Fig.\ 
\ref{fig:lambdas_vs_u}).}
\label{fig:susc_and_tks}
\end{center}
\end{figure*}

\section{Effective gating due to capacitive coupling}
 Results from the variational method are shown in Figure 
\ref{fig:lambdas_vs_u}(a) as function of the capacitive coupling $u$, for 
impurities with identical couplings to their respective metallic leads. 
While $\varepsilon_d \ll -\Gamma_d$ is fixed deep in the Kondo regime, and 
$\varepsilon_q (\ge\varepsilon_d)$  is set closer to the mixed valence 
regime ($\varepsilon_q \lesssim -\Gamma_q$), we see that increasing $u$ 
affects both subsystems asymmetrically. As $u$ increases, the level of 
impurity $q$ is raised ($\Lambda_q$ increases), while the remote gate effect due 
to the capacitive coupling vanishes for impurity $d$ ($\Lambda_d$ 
decreases). The remote gate then becomes effectively larger for impurity 
$q$, indicating that the ``shallower'' level is more susceptible to the 
capacitive coupling. One can qualitatively understand this behavior by 
examining how a shift in $\varepsilon_q$ affects the Kondo temperature. 
Substituting $\varepsilon_q \rightarrow \varepsilon_q + \Lambda_q$ into the 
expression for the Kondo temperature within the variational 
framework\cite{varma_yafet_1976}
\begin{equation}\label{eq:vonDelft}
T_K^i = D\exp\left( -\pi \abs{\varepsilon_i}/\Gamma_i \right),
\end{equation}
and evaluating its total differential as a function of $\Lambda_q$, we 
find
\begin{equation}\label{eq:haldane_shift}
\ud T_{K}^{q} \equiv \frac{\partial T_K^q}{\partial \varepsilon_q} \ud 
\varepsilon_q =  -\Lambda_q \,\frac{\text{sgn}\left[ \varepsilon_q 
\right] \,\pi D}{\Gamma_q}\,\exp( -\pi|\varepsilon_q|/\Gamma_q ),
\end{equation}
 which is proportional to $\Lambda_q$ and grows exponentially as 
$\abs{\varepsilon_q}$ becomes smaller.\footnote{While the variational 
approach corresponds to the case of infinite intra--dot Coulomb 
interaction, the variation of the ground--state energy can be generalized 
with the use of Haldane's formula\cite{Haldane_1978}, to $\ud T_K^i \sim 
\Lambda_i(U_i\Gamma_i)^{-1/2}(2\varepsilon_i + U_i)\exp\{ 
\pi\varepsilon_i(\varepsilon_i + U_i)/(2U_i\Gamma_i) \}$.} Notice that the 
Kondo temperature is a measure of how much the impurity hybridization contributes to the 
lowering of the ground state energy of each subsystem due to the onset of 
many-body correlations, with respect to the atomic limit ($\Gamma_q = 0$). 
The essence of Eq.\ (\ref{eq:haldane_shift}) is that it is energetically 
favorable for the ``shallower'' level to be shifted the most when the 
coupling increases. A maximum shift of $\Lambda_q \approx u$ for large values 
of $u$ is in accordance with Eqs.\ (\ref{eq:u_and_lambdad}) and 
(\ref{eq:u_and_lambdaq}).

\begin{figure}
\begin{center}
\includegraphics[scale=0.5]{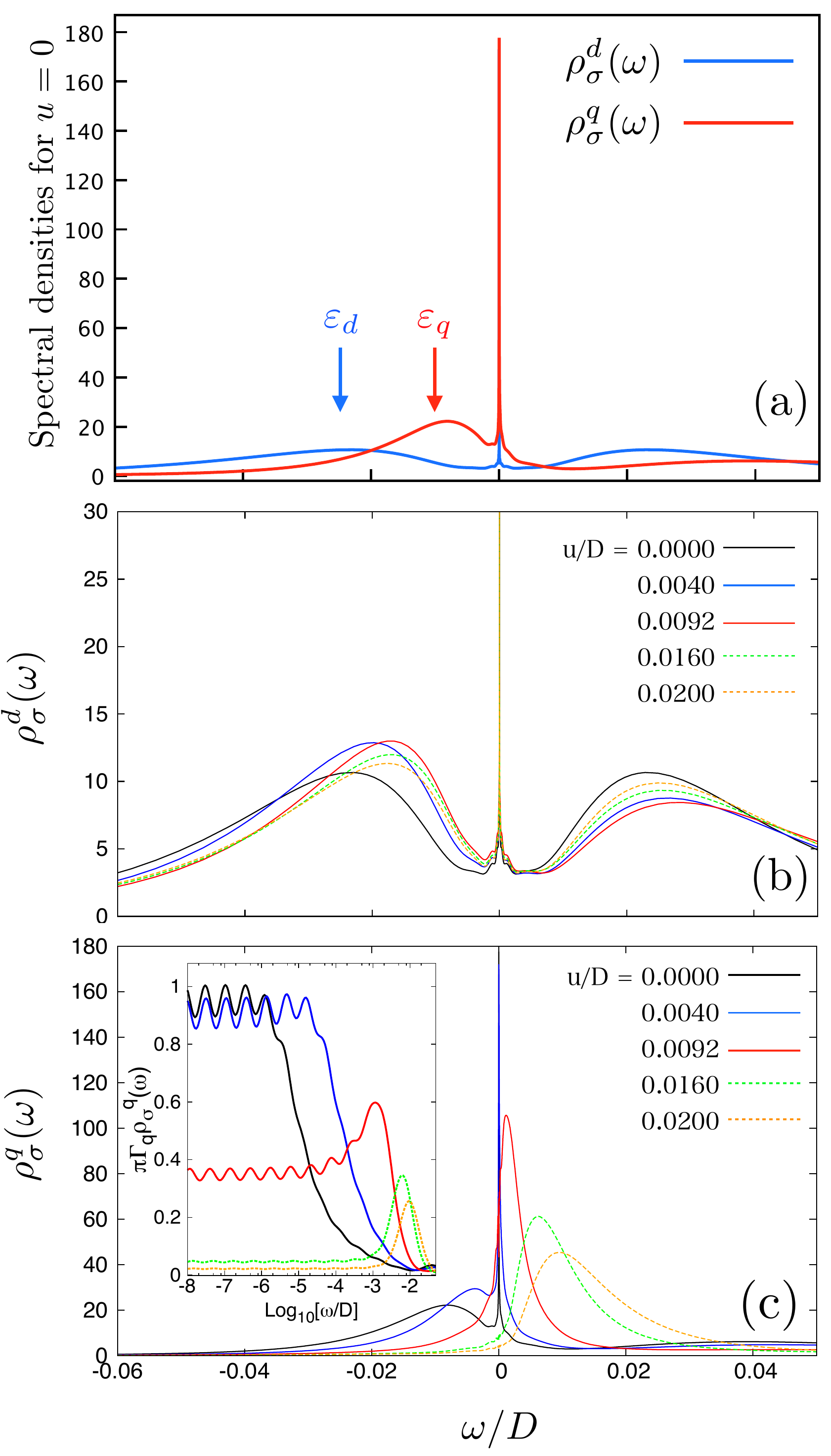}
\caption{(Color online) Spectral densities of both impurities for varying 
capacitive coupling $u$, for parameters $U_d = U_q = 0.05\,D$, 
$\varepsilon_d = -U_d/2 = -0.025\,D$, $\varepsilon_q = -0.010\,D$ and 
$\Gamma_d = \Gamma_q = 0.002\,D$. (a) Reference curves with $u/D = 0$, 
showing the local levels dressed by their respective electronic reservoirs, and the ASR of each impurity at the Fermi 
level. Same $u=0$ curves shown in black in corresponding (b) and (c) 
panels. (b) The ASR of impurity $d$ remains well--defined for all values of 
$u$. The panel focuses on the Hubbard bands, amplifying the lower region of the 
vertical axis for clarity (notice different scale). 
(c) In the case of impurity $q$, the disappearance of the ASR and 
the shift of the left Hubbard peak toward positive values of $\omega$ with 
increasing $u$ is a clear indication that the Kondo effect is being 
quenched. Inset in (c) shows detailed behavior of 
$\rho_{\sigma}^{q}(\omega)$ for $\omega \approx 0$. For $u \gtrsim 
0.01\,D$, the $q$-impurity is pushed out of the Kondo regime.}
\label{fig:spectral_vs_u}
\end{center}
\end{figure}

A comparison between the variational and NRG approaches is shown in Fig.\ 
\ref{fig:lambdas_vs_u}(b), where the charges and charge fluctuations of 
both impurities are presented. Impurity $q$ is depleted due to the external 
gating produced by the capacitive coupling, in both calculations. 
Although it is clear from comparing to the NRG results that the variational 
method overestimates the strength of the gating---likely a consequence of 
the infinite-$U_i$ approximation that allows the simple form of the proposed ground 
state---the behavior predicted by both methods is qualitatively 
consistent.  As $u$ increases, $\langle n_q \rangle$ decreases to zero,
while $\langle n_d \rangle$ remains essentially constant $\simeq 1$.  
Correspondingly, the characteristic fluctuations in $q$ are significant and
peak (at $\langle n_q \rangle \simeq \frac{1}{2}$), whereas those in $d$
remain small throughout.

 Turning to the thermodynamic properties of the system, Figs.\ 
\ref{fig:susc_and_tks}(a) and (c) show the temperature dependence of the 
effective magnetic--moment--squared of impurities $d$ and $q$, 
respectively, given by $\mu_i(T) = T\,\chi_i(T)$, with $\chi_i$ the 
magnetic susceptibility of impurity $i$. Different curves correspond to 
different values of the capacitive coupling. The Kondo temperature in each 
case can be obtained from each curve [$T_K^i \chi(T_K^i) = 0.0707$], and 
the temperature axis can be rescaled as in Figures 
\ref{fig:susc_and_tks}(b) and (d), in order to collapse all curves into the 
universal curve that is the hallmark of Kondo physics.\cite{wilson_1975} 
All curves for the $d$-impurity demonstrate this universal behavior in 
Fig.\ \ref{fig:susc_and_tks}(b). Notice, however, that impurity--$q$ curves 
for $u/D > 0.0092$ fall outside the universality curve in Fig.\ 
\ref{fig:susc_and_tks}(d), indicating a transition away from the Kondo 
regime for large $u$, which from the variational analysis can be understood as 
the large gating coming from the capacitive coupling. This is consistent 
with the depletion of impurity $q$ shown in Fig.\ 
\ref{fig:lambdas_vs_u}(b). The exponential dependence of the Kondo 
temperature $T_K^q$ on $u$ [see Eq.\ (\ref{eq:vonDelft})], shown in the 
inset of Fig.\ \ref{fig:susc_and_tks}(c), is further confirmation of this 
picture. It is important to emphasize that increasing the gating on 
impurity $q$ at first enhances its Kondo temperature. However, beyond a 
critical value of $u$, the impurity reaches the mixed--valence regime, and 
is eventually emptied out. In contrast, as the shallower impurity is 
depleted by the gating, the Kondo temperature $T_K^d$ of impurity $d$ is in 
fact restored to its uncoupled value, as the inset of Fig.\ 
\ref{fig:susc_and_tks}(c) shows.

To complete our analysis of this case, we present the local spectral 
density of the impurities in Fig.\ \ref{fig:spectral_vs_u}. The spectral 
density of impurity $q$ is defined as
\begin{equation}
\rho_{\sigma}^{q}(\omega)=-\frac{1}{\pi}\text{Im}\gf{q_{\sigma}}{q_{\sigma}
^{\dagger}}{\omega},
\end{equation}
with $\gf{A}{B}{\omega}$ the retarded Green's function\cite{zubarev_1960} 
of operators $A$ and $B$. It represents the effective single--particle 
level density available at energy $\omega$, at the impurity site: the left 
peak at $\omega/D \approx -0.01$ of the curve for $u/D = 0$ in Fig.\ 
\ref{fig:spectral_vs_u}(a) (solid red line) corresponds to the occupied 
impurity level, dressed by the leads' electrons. The sharp peak at 
$\omega/D = 0$ is the Abrikosov--Suhl resonance (ASR)--- a typical signature 
of the Kondo effect.\cite{abrikosov_1965,suhl_1965,nagaoka_1965} As the 
capacitive coupling increases, there is a clear progression of the 
effective $q$-level toward positive values, as seen in Fig.\ 
\ref{fig:spectral_vs_u}(c), eventually quenching the Kondo effect at $u/D 
\approx 0.0092\,D$, when the charge fluctuations peak, and the charge is reduced by
 50\% (see Fig.\ \ref{fig:lambdas_vs_u}). In fact, assuming a simple 
linear shift of $\varepsilon_q \rightarrow \varepsilon_q + u$ would suggest that 
the mixed--valence regime will be reached for $u \approx |\varepsilon_q| - 
\Gamma_q$, in agreement with the variational result [Fig.\ 
\ref{fig:lambdas_vs_u}(a)] that yields $\Lambda_q \rightarrow u$ for $u 
\rightarrow |\varepsilon_q|$. That the Kondo effect is quenched for slightly
larger $u$ reflects the overestimation of the remote gating by the variational
method, as mentioned above. Figure \ref{fig:spectral_vs_u}(c) also shows 
that for $u \gtrsim 0.0092\,D$, the ASR at the Fermi level disappears in the 
spectral function for the $q$-impurity, and the (now emptying) 
single--particle resonance moves increasingly above the Fermi level. The 
inset in Fig.\ \ref{fig:spectral_vs_u}(c) examines the behavior of the 
$\rho^{q}_{\sigma}$ near the Fermi level, showing again its decreasing 
value, vanishing for $u \gtrsim 0.0092\,D$ as well. The spectral density 
$\rho_{\sigma}^{d}(\omega)$, on the other hand, shows only a slight, 
non--monotonic shift of the impurity--$d$ level, with a stable ASR for all 
values of the capacitive coupling, also in agreement with the variational 
result [see Fig.\ \ref{fig:spectral_vs_u}(b)].

\section{The role of particle--hole symmetry in the coupled system}
\begin{figure}
\begin{center}
\includegraphics[scale=0.42]{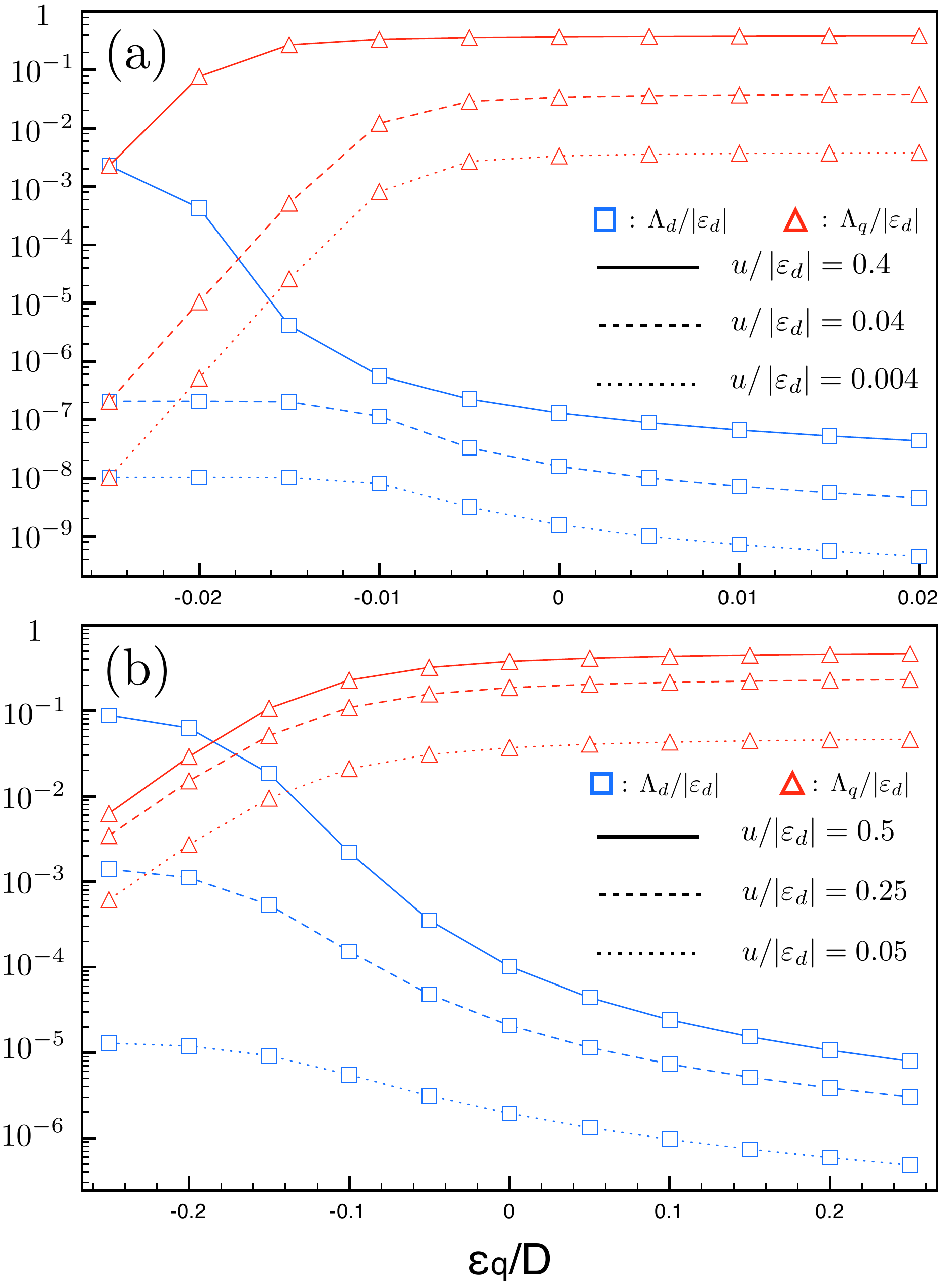}
\caption{(Color online) External gating of the impurity levels, 
$\Lambda_d$ and $\Lambda_q$ as a function of $\varepsilon_q$, as
obtained from the 
variational calculation. (a) For impurities with identical couplings 
($\Gamma_d = \Gamma_q = 0.002\,D$), with fixed $\varepsilon_d = -0.025\,D$. 
(b) For different couplings ($\Gamma_d=0.002\,D$, $\Gamma_q = 0.05\,D$), 
with fixed $\varepsilon_d = -0.02\,D$.}
\label{fig:lambdas_vs_Eq}
\end{center}
\end{figure}

Armed now with the basic understanding of the inter--impurity coupling 
model, we turn to a setup that is commonly implemented in experiments. 
It is possible to independently tune the impurity levels in a QD sample by 
means of gate voltages---a common practice in electron counters for 
QDs\cite{gustavsson_prl_2006} and ``which path'' 
interferometers.\cite{aleiner_prl_1997,buks_nature_1998,
avinun-kalish_prl_2004}  We study now the coupled system when
one of the quantum impurity levels is controlled by an external gate.

Considering again the case $\Gamma_d=\Gamma_q$, Fig.\ 
\ref{fig:lambdas_vs_Eq}(a) shows results of the variational method for 
different fixed values of $u$, while varying $\varepsilon_q$ but keeping 
$\varepsilon_d = -0.025\,D$ fixed. Starting at $\varepsilon_q = 
\varepsilon_d$, both impurities experience the same gating $\Lambda_d = 
\Lambda_q$, as one would expect from symmetry arguments. As the level of 
impurity $q$ moves closer to the Fermi energy, $\Lambda_q$ grows quickly, 
while $\Lambda_d$ decreases. Once again, it is clear that the shallower 
level--- that of impurity $q$--- experiences a larger gating. Notice that 
the asymmetry persists even for small $u$, where $\Lambda_q$ grows fast, 
while $\Lambda_d$ drops only slightly. This general behavior of the 
$\Lambda_i$ persists for impurities with different lead couplings, as can 
be seen in Fig.\ \ref{fig:lambdas_vs_Eq}(b), which shows results for 
$\Gamma_q = 0.05\,D$, $\Gamma_d = 0.002\,D$ and $\varepsilon_d = -0.02\,D$. 
The asymmetry of the subsystems is reflected in the different 
values of $\Lambda_d$ and $\Lambda_q$, even for $\varepsilon_q \le 
\varepsilon_d$.

\begin{figure}%[hbt!]
\begin{center}
\includegraphics[scale=0.44]{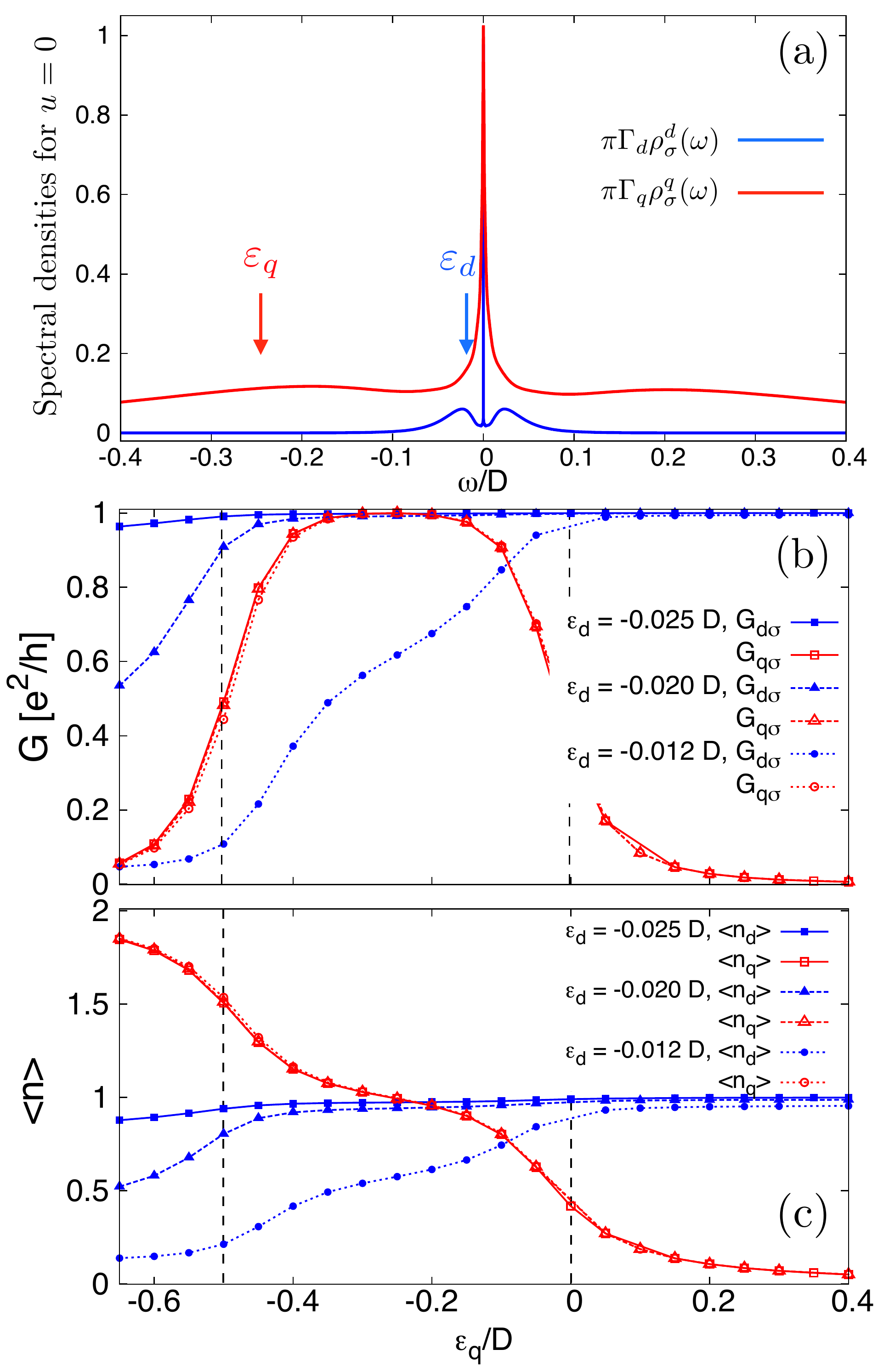}
\caption{(Color online) (a) Comparison of the spectral densities of  both 
impurities (in the electron--hole symmetric regime, 
$\varepsilon_d=-U_d/2$), when $\Gamma_q \gg \Gamma_d$ and $U_q \gg U_d$. 
(b) Conductances per spin channel (in units of $e^2/h$) for impurities $d$ 
and $q$, and (c) their occupations, as functions of $\varepsilon_q$, for 
different values of  $\varepsilon_d$. All results from NRG calculations; 
parameters: $U_d=0.05\,D$, $U_q = 0.5\,D$, $\Gamma_d = 0.002\,D$, $\Gamma_q 
= 0.05\,D$, $u=0.01\,D$. Vertical dashed lines denote the region in the
uncoupled regime ($u=0$) for impurity $q$ ($\langle n_q \rangle=1$) where
the high conductance in the Kondo regime is expected.}
\label{fig:condd_vs_Eq}
\end{center}
\end{figure}

\begin{figure}%[hbt!]
\begin{center}
\includegraphics[scale=0.30]{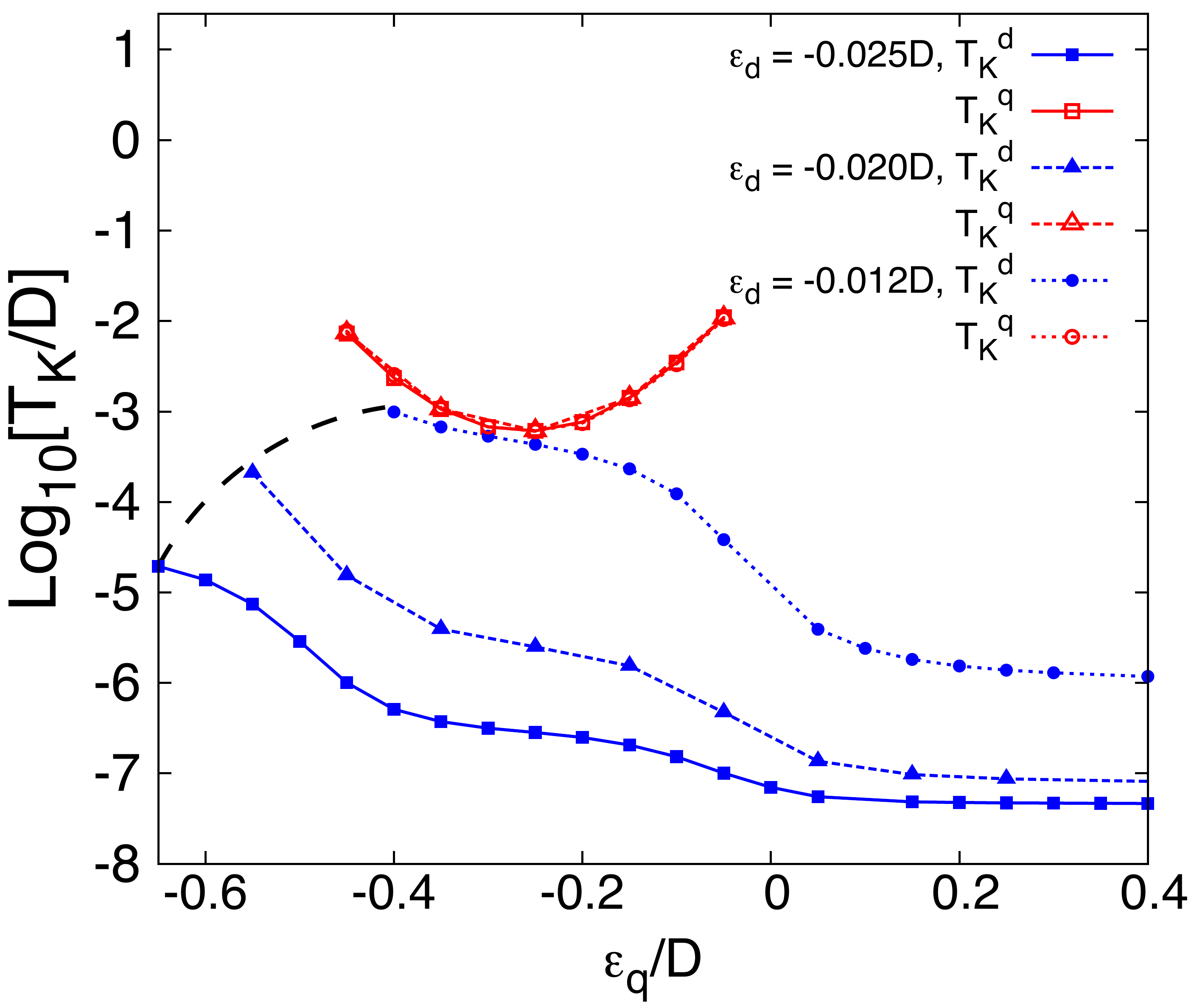}
\caption{(Color online) Kondo temperatures $T_K^d$ and $T_K^q$ of 
impurities $d$ and $q$, respectively, as functions of $\varepsilon_q$, for 
three fixed values of $\varepsilon_d$. The curves terminate, indicating the 
approximate values of $\varepsilon_q$ at which the Kondo effect is quenched 
in each case. The dashed line on the left indicates the trend of the 
transition out of the Kondo regime for impurity $d$. All other parameters 
as in Fig.\ \ref{fig:condd_vs_Eq}.}
\label{fig:TKs_vs_Eq}
\end{center}
\end{figure}

Naturally, in a QD system one can monitor the different ground states via 
conductance measurements. In Fig.\ \ref{fig:condd_vs_Eq} we present NRG 
calculations of the zero--bias conductance through each impurity, with 
$\varepsilon_d$ fixed while $\varepsilon_q$ is varied. The conductance 
profile of impurity $q$ remains approximately the same for all three values 
of $\varepsilon_d$ shown; it is, in fact, nearly identical to the case of a 
completely independent impurity ($u=0$), since it is so deep into the Kondo
regime and the interaction $u$ does not affect it much. 
Impurity $d$--- here the shallower 
one---, on the other hand, is strongly affected by the capacitive 
coupling, especially for $\varepsilon_q < -U_q/2$, when $\xpect{n_{q}} > 1$ 
(Fig.\ \ref{fig:condd_vs_Eq}). While its bare parameters are within the 
Kondo regime, the enhanced conductance of impurity $d$--- characteristic of 
the Kondo effect--- is strongly suppressed as the capacitive coupling to 
$q$ shifts $\varepsilon_d$ effectively, and reduces its average charge 
[Fig.\ \ref{fig:condd_vs_Eq}(c)]. In accordance with the Friedel sum 
rule,\cite{hewson_kpthf} the conductance per spin channel at zero 
temperature is given by 
\begin{equation}\label{eq:friedel}
G_{d\sigma}(T=0) = \frac{e^2}{h}\,\sin\left(\pi \xpect{n_{d\sigma}} 
\right),
\end{equation}
and so it will be reduced as $\xpect{n_{d\sigma}} < 1$.  The conductance 
curves in Fig.\ \ref{fig:condd_vs_Eq}(b) reflect the variation of 
$\xpect{n_{d\sigma}}$ as the gate applied to impurity $q$ shifts its level 
[Fig.\ \ref{fig:condd_vs_Eq}(b)], and it shows how the enhanced conductance 
may be reduced as a consequence of the \emph{equilibrium charge 
fluctuations} in the impurity. This effect is naturally more pronounced for 
$\varepsilon_d$ closer to the Fermi level, yet well into the Kondo regime 
for $u=0$.
The influence of the remote gating from impurity $q$ on impurity $d$ is 
perhaps more clearly appreciated in Fig.\ \ref{fig:TKs_vs_Eq}, which shows 
the Kondo temperatures of both impurities as functions of $\varepsilon_q$. 
The termination of each of the curves indicate the (approximate) values of 
$\varepsilon_q$ at which the Kondo effect is quenched in either impurity. 
As in the conductance results, the curves for impurity $q$ are nearly 
identical to the independent impurity limit. The curves of $T_K^d$, on the 
other hand, show a plateau structure that directly relates to 
$\xpect{n_{q\sigma}}$ [see Fig.\ \ref{fig:condd_vs_Eq}(c)]: As 
$\varepsilon_q$ grows more negative, the remote gating on $d$ increases 
with $u\xpect{n_{q}}$---the energy contribution of the capacitive 
coupling---raising the level of impurity $d$, and consequently increasing 
$T_K^d$ until the Kondo effect is quenched. This transition is indicated 
by the black, dashed trend line in the figure. As the remote gating 
decreases (with increasing $\varepsilon_q$) and the enhanced conductance is 
restored in $d$, the Kondo temperature $T_K^d$ is reduced back to its value 
in the electron--hole symmetric case. At this point, it is impurity $q$ 
that is gated the most, but given that $|\varepsilon_q| \gg u$, the effects 
of this gating are not noticeable in either the conductance or the Kondo 
temperatures. Notice that, as the $q$-impurity is emptied 
($\varepsilon_q >0$), $T_K^{d}$ settles into the isolated $d$-impurity 
value.

\begin{figure}%[hbt!]
\begin{center}
\includegraphics[scale=0.36]{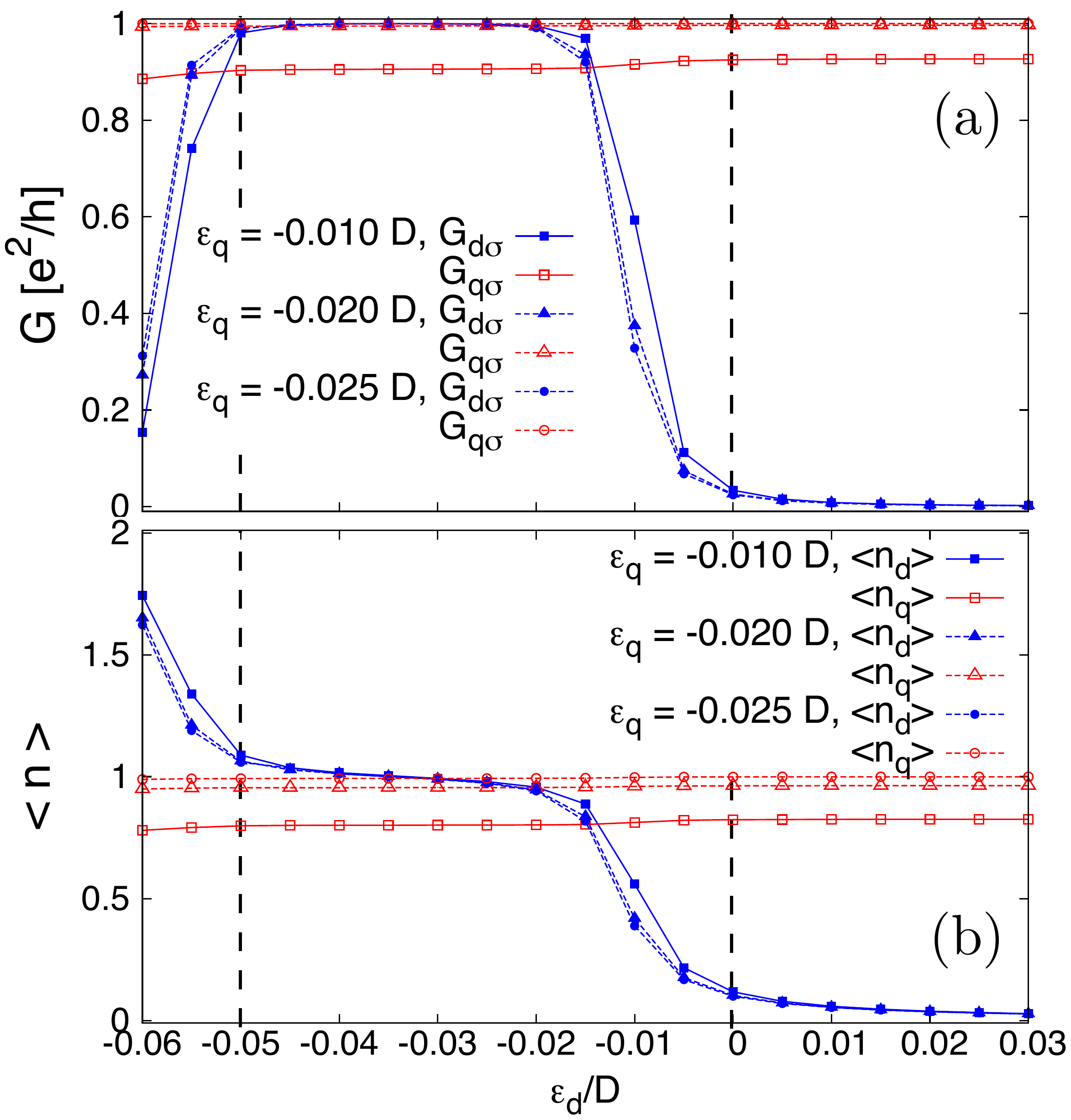}
\caption{(Color online) NRG calculations of (a) the conductance of 
impurities $d$ and $q$, and (b) the occupation of both impurities, as 
functions of $\varepsilon_d$, for different values of  $\varepsilon_q$. 
Parameters: $U_d=0.05\,D$, $U_q = 0.5\,D$, $\Gamma_d = 0.002\,D$, $\Gamma_q 
= 0.05\,D$, $u=0.01\,D$. Vertical dashed lines indicate the region 
$\langle n_d \rangle =1$ for the uncoupled case ($u=0$).  Notice
the typical conductance plateau is shifted by $\approx u$ to lower $\varepsilon_d$ 
values due to the remote gating effect.}
\label{fig:condq_vs_Ed}
\end{center}
\end{figure}

The same conductance analysis is repeated in Fig.\ \ref{fig:condq_vs_Ed}, 
this time varying the level of impurity $d$, which has the smaller 
parameters $U_d$ and $\Gamma_d$. The gating effects on impurity $q$ are 
negligible, with only a slight drop in conductance (about $10\%$) in the case of 
$\varepsilon_q=-0.010\,D$, when it is on the edge between the Kondo phase 
and the mixed--valence regime. Most noticeable is the overall shift of 
$\approx u$ on the conductance profile of impurity $d$ to lower 
$\varepsilon_d$ values, as a result of the gating.

\begin{figure*}%[hbt!]
\begin{center}
\includegraphics[scale=0.55]{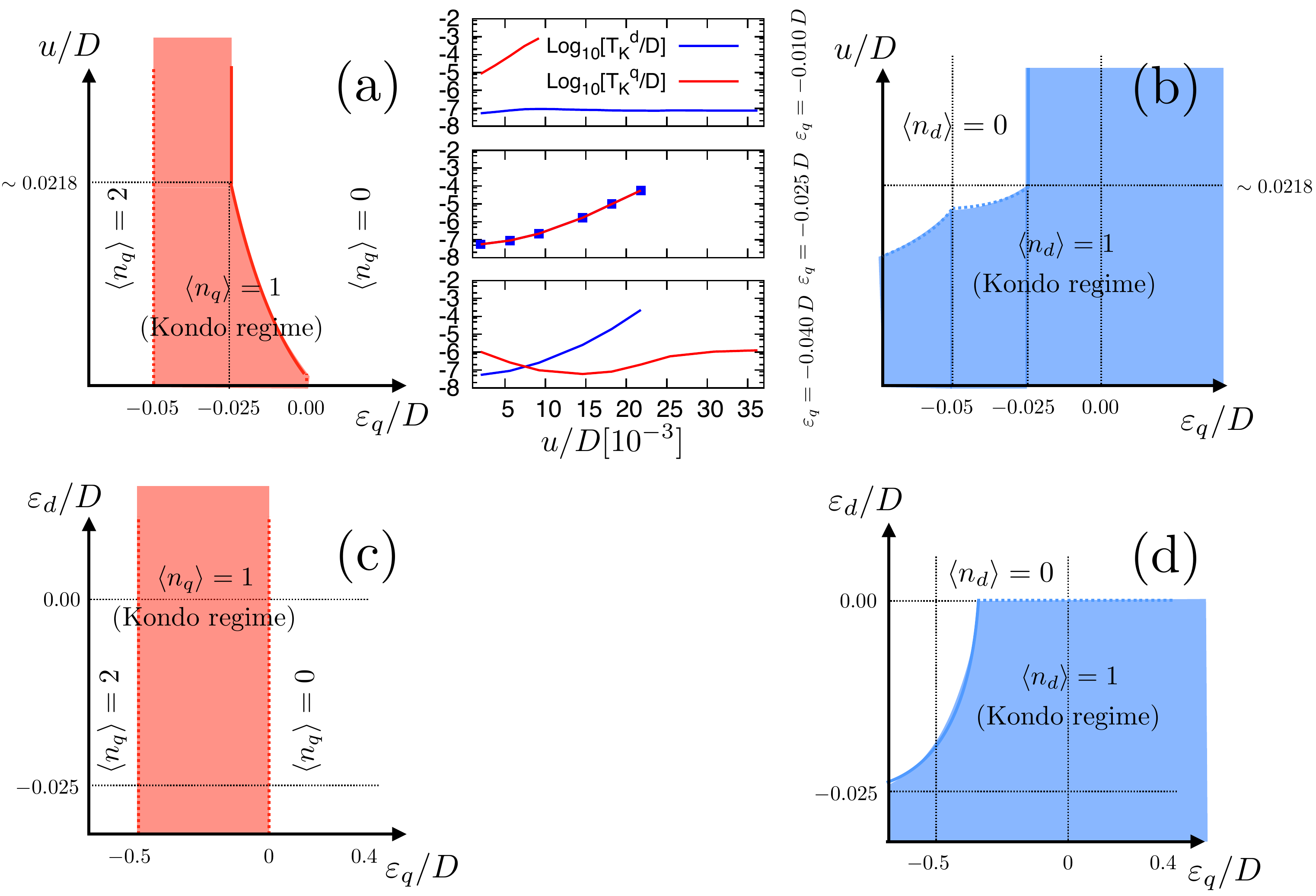}
\caption{(Color online) Schematic $u$-$\varepsilon_q$ phase diagrams for (a) impurity $q$ and 
(b) impurity $d$, for fixed $\varepsilon_d = -0.025\,D$, with $U_d = U_q = 
0.05\,D$ and $\Gamma_d = \Gamma_q = 0.002\,D$. The Kondo temperatures of 
both impurities for select values of $\varepsilon_q$ are shown in the 
central panels, as functions of $u$. (c) and (d) show the phase diagrams
in the $\varepsilon_d$-$\varepsilon_q$ plane for 
impurities $q$ and $d$, respectively, for fixed $u = 0.01\,D$, with $U_d = 
0.05\,D$, $U_q = 0.5\,D$, $\Gamma_d=0.002\,D$ and $\Gamma_q=0.05\,D$. The 
shaded areas indicate that the corresponding impurity is in the Kondo 
regime. The Kondo temperatures are higher closer to the boundaries of the 
shaded regions.}
\label{fig:phase_diagrams}
\end{center}
\end{figure*}

The results presented in this section are summarized in Fig.\ 
\ref{fig:phase_diagrams}, which shows the phase diagrams for both 
subsystems. Figure \ref{fig:phase_diagrams}(a) shows the quenching of the 
Kondo effect in impurity $q$ by the capacitive coupling strength $u$ for 
fixed $\varepsilon_d$, as a transition from $\xpect{n_q} = 1$ to $0$, for 
any given value of $\varepsilon_q$ within the colored region. Fig.\ 
\ref{fig:phase_diagrams}(b) demonstrates how the $d$-impurity undergoes 
this transition only for $\varepsilon_q < \varepsilon_d$, when it becomes 
the shallower level. The two slopes of the left boundary of the colored 
region indicate different behaviors of impurity $d$ when $\xpect{n_q}=1$ 
and $2$. The location of these boundaries are inferred from the 
behavior of the Kondo temperatures as the structure parameters change.
$T_K^d$ and $T_K^q$ are shown in the middle panel of the figure for
typical cases.  In the upper panel, $q$ is the shallower subsystem 
($|\varepsilon_q|/\Gamma_q \ll |\varepsilon_d|/\Gamma_d$), and as $u$ increases
it is $q$ which is effectively gated into the empty impurity regime.  The middle
panel shows a fully symmetric case, where large $u$ destroys the Kondo
regime in both subsystems simultaneously.  Finally, the bottom panel shows
the case where $d$ is shallower ($|\varepsilon_q|/\Gamma_q \gg |\varepsilon_d|/\Gamma_d$)
such that the gating destroys the Kondo correlations in $d$.  In the asymmetric
cases, it is also clear that as the other subsystem is gated into the empty regime, 
the Kondo temperature of the remaining impurity reverts to the isolated value
for increasing $u$. 

Figures \ref{fig:phase_diagrams}(c) and (d) illustrate the phase diagram 
for fixed $u$, while varying $\varepsilon_d$ and $\varepsilon_q$. Because 
$|\varepsilon_q|$ is very large with respect to $u$, the transition out of 
Kondo is induced only on impurity $d$, whose level depth $\varepsilon_d$ is 
comparable to the capacitive coupling strength $u$. This transition, induced 
by the presence of impurity $q$, is shown as the left boundary to the blue 
region in panel (c), and occurs for lower values of $\varepsilon_d$ as 
$\xpect{n_q}$ grows (for more negative values of $\varepsilon_q$) and the 
capacitive interaction strength increases.

These results are important and should be considered in the analysis of coupled
systems, such as the decoherence 
studies where the suppression of the 
zero--bias conductance through the QD is used to measure the dephasing rate 
of a nearby charge detector.\cite{avinun-kalish_prl_2004}  As shown in Fig.\ \ref{fig:condd_vs_Eq}, the 
capacitive coupling between the two impurities in equilibrium can 
significantly suppress the enhanced conductance expected of the quantum 
dot, by virtue of the competition it introduces between the impurities' 
ground states. Charge detectors for quantum dots are commonly implemented 
using a quantum point contact, and it has been suggested\cite{aono_2008} 
that the anomalous behavior observed in the dephasing rate of the QPC in 
the experiments of Ref.\ [\onlinecite{avinun-kalish_prl_2004}] is due to 
the appearance of a localized level within the QPC, which can undergo Kondo 
screening.\cite{meir_prl_2002} Our study is thus highly relevant for the 
interpretation and full analysis of these and similar experiments, 
as it presents an equilibrium analysis of the experimental setup, which 
is typically treated as a static system (the limit of Fig.\ 
\ref{fig:condq_vs_Ed}). We demonstrate that, in some parameter regimes, the 
ground state of the system is in fact quite sensitive to capacitive 
coupling, and the equilibrium charge fluctuations of one of the impurity 
subsystems are able to strongly influence the conductance of the other.
This sensitivity makes it a much  more subtle task to evaluate how much of the conductance 
suppression in one subsystem is due to the dephasing induced by the other.

\section{Conclusions}

We have studied the Kondo physics of a system of two capacitively--coupled 
quantum impurities, by means of both a variational method and numerical 
renormalization group calculations. We found that the capacitive 
interaction can be absorbed into an external gating effect that varies 
non--linearly with the capacitive coupling strength, as well as the 
impurities' energy levels. We demonstrate that the external gating to one 
impurity coming from the other can induce a quantum phase transition out of 
the Kondo regime, modifying its conductance properties in a way that can be 
measured experimentally. Our results suggest that, as the subtle balance of 
interactions strongly affects the nature of this type of system in certain 
parameter regimes, it is of great relevance to the study of dephasing 
effects of charge detectors on quantum dots, where conductance suppression 
is used as a measure of the dephasing rate.

\appendix
\section{Variational calculation of the two--impurity Kondo ground 
state}\label{app:variational}
We use a variational approach following the methods of Ref.\ 
[\onlinecite{varma_yafet_1976}], where a variational wavefunction for the 
ground state is proposed based on the many--body singlet nature of the 
Kondo state. The $d$-impurity ground state is given by
\begin{equation}\label{eq:QDgnd}
\ket{\psi_{d}} = \left[\beta_0+ \sum_{\mathbf{k}\sigma} 
d_{\sigma}^{\dagger} \beta_{\mathbf{k}}\,c_{d\mathbf{ 
k}\sigma}\right]\ket{\Omega_{d}},
\end{equation}
with the state
\begin{equation}
\ket{\Omega_{d}} = \prod_{\abs{\mathbf{k}}<k_F}{c_{d\mathbf{k} 
\uparrow}}^{\dagger}{c_{d\mathbf {k}\downarrow}}^{\dagger}\ket{0},
\end{equation}
representing the leads filled up to the Fermi momentum $k_F$. The 
corresponding state for impurity $q$ is
\begin{equation}\label{eq:QPCgnd}
\ket{\psi_{q}} = \left[\alpha_0 + \sum_{\mathbf{k}\sigma} 
q_{\sigma}^{\dagger} \alpha_{\mathbf{k}}\,c_{q\mathbf 
{k}\sigma}\right]\ket{\Omega_{q}}.
\end{equation}
The set of (real) coefficients $\alpha = \{\alpha_0,\,\alpha_{\mathbf{k}} 
\}$ and $\beta=\{\beta_0,\,\beta_{\mathbf{k}} \}$ are variational 
parameters. These variational states are expected to be more reliable in 
the limit of large $U_i$, as double occupancy is omitted. The full state 
of 
the coupled system is then proposed as
\begin{equation}\label{eq:gnd}
\ket{\Psi_{\text{gnd}}} = \ket{\psi_{q}}\otimes\ket{\psi_{d}},
\end{equation}
and the energy contribution from the impurities is given by
\begin{equation}\label{eq:totalenergy}
E(\alpha,\,\beta) = \frac{\expect{\Psi_{\text{gnd}}} 
{H}{\Psi_{\text{gnd}}}}{\braket{\Psi_{\text {gnd}}}{\Psi_{\text{gnd}}}} = 
E_{d} + E_{q} + \tilde{u},
\end{equation}
where we have defined the separate impurity energy contributions
\begin{equation}
E_{q} = 2\xpect{\psi_{q}}^{-1}\sum_{\abs{\mathbf{k}}<k_F} 
\left[\left(\varepsilon_{ q} - \varepsilon_{\mathbf{k}}^q  
\right)\alpha_{\mathbf{k}}^2 - 2V_q\alpha_0\alpha_{\mathbf{k}} \right],
\end{equation}
\begin{equation}
E_{d} =  2\xpect{\psi_{d}}^{-1}\sum_{\abs{\mathbf{k}}< 
k_F}\left[\left(\varepsilon_{ d} - \varepsilon_{\mathbf{k}}^{d} 
\right)\beta_k^2 - 2V_d\beta_0\beta_{\mathbf{k}} \right],
\end{equation}
and the capacitive coupling contribution
\begin{equation}
\tilde{u} = 
u\,\frac{\left(2\sum_{\abs{\mathbf{k}}<k_F}\alpha_{\mathbf{k}}
^2\right)\left(2\sum_{\abs{\mathbf{k}}<k_F}\beta_{\mathbf{k}}^2\right)}{
\braket{\psi_{d}}{\psi_{d}}\braket{\psi_{q}}{\psi_{q}}}.
\end{equation}
We then proceed to minimize the ground state energy Eq.\ 
(\ref{eq:totalenergy}) with respect to the variational amplitudes by the 
conditions
\begin{subequations}
\begin{equation}
\frac{\partial}{\partial \alpha_0}E(\alpha,\,\beta) = 
\frac{\partial}{\partial \alpha_k}E(\alpha,\,\beta) = 0,
\end{equation}
\begin{equation}
\frac{\partial}{\partial \beta_0}E(\alpha,\,\beta) = 
\frac{\partial}{\partial \beta_k}E(\alpha,\,\beta) = 0.
\end{equation}
\end{subequations}
The resulting variational equations are
\begin{subequations}\label{eq:vareqq}
\begin{equation}\label{eq:vareqq1}
2V_q\sum_{\abs{\mathbf{k}}<k_F}\alpha_{\mathbf{k}} = \alpha_0\left( E_q + 
\tilde{u} \right),
\end{equation}

\begin{equation}\label{eq:vareqq2}
\begin{split}
\alpha_0V_q= \alpha_{\mathbf{k}}\Big[ &E_{q} - \left(\varepsilon_{q} - 
\varepsilon_{\mathbf{k}}^{q} \right) \\ &- 
2u\,\alpha_0^2\braket{\Psi_{\text{gnd}}} 
{\Psi_{\text{gnd}}}^{-1}\sum_{\abs{ \mathbf{k}}<k_F}\beta_{\mathbf{k}}^2  
\Big],
\end{split}
\end{equation}
\end{subequations}
and a completely identical set for $d$ (\emph{i.e.}, $q \rightarrow d$ and 
$\alpha \rightarrow \beta$). We propose the relations $\alpha_{\mathbf{k}} 
\equiv \alpha_k$ and $\beta_{\mathbf{k}} \equiv \beta_k$, with
\begin{subequations}
\begin{equation}
\alpha_k = \frac{\alpha_0V_q}{\varepsilon_k^{q} - \varepsilon_{q} - 
\Lambda_{q} + E_{q}},
\end{equation}

\begin{equation}
\beta_k = \frac{\beta_0V_d}{\varepsilon_k^{d} - \varepsilon_{d} - 
\Lambda_{d} + E_{d}},
\end{equation}
\end{subequations}
and substitute into Eqs.\ (\ref{eq:vareqq}) and the corresponding 
equations for impurity $d$ to obtain (in the continuum limit)
\begin{subequations}\label{eq:sysofeqs}
\begin{equation}
\begin{split}
\frac{V_d^2}{D}\log\left[\frac{\varepsilon_{q} - E_{d} -\varepsilon_F + 
\Lambda_{q}}{\varepsilon_{q} - E_{q} + D + \Lambda_{q}} \right] =&\\
E_{q} + u\frac{X_qX_d}{\left(1+X_q \right)\left(1+X_d \right)},&
\end{split}
\end{equation}
\begin{equation}
\begin{split}
\frac{V_q^2}{D}\log\left[\frac{\varepsilon_{d} - E_{q} -\varepsilon_F + 
\Lambda_{d}}{\varepsilon_{d} - E_{d} + D + \Lambda_{d}} \right] =&\\
E_{d} + u\frac{X_qX_d}{\left(1+X_q \right)\left(1+X_d \right)},&
\end{split}
\end{equation}
\end{subequations}

\begin{subequations}
\begin{equation}\label{eq:u_and_lambdad}
\Lambda_d = u\frac{X_q}{\big(1+X_d \big)\big(1+X_q \big)},
\end{equation}
\begin{equation}\label{eq:u_and_lambdaq}
\Lambda_q = u\frac{X_d}{\big(1+X_d \big)\big(1+X_q \big)}.
\end{equation}
\end{subequations}
where we have defined the continuum--limit quantities
\begin{subequations}
\begin{equation}\label{eq:chq_to_lambdaq}
\alpha_0X_q = 2\sum_{k<k_F}\alpha_k^2 \rightarrow 
\frac{V_q^2}{\left(\varepsilon_q - E_q + \Lambda_q \right) 
\left(\varepsilon_q - E_q + D + \Lambda_q  \right)},
\end{equation}
\begin{equation}\label{eq:chd_to_lambdad}
\beta_0X_d = 2\sum_{k<k_F}\beta_k^2 \rightarrow 
\frac{V_d^2}{\left(\varepsilon_d - E_d + \Lambda_d \right) 
\left(\varepsilon_d - E_d + D + \Lambda_d  \right)}.
\end{equation}
\end{subequations}
This system of equations is then solved numerically, in terms of 
$\Lambda_d$ and $\Lambda_q$. From these solutions, and using Eqs.\ 
(\ref{eq:chq_to_lambdaq}) and (\ref{eq:chd_to_lambdad}), we directly 
obtain 
the charge and charge fluctuations of each impurity.

\section{The conductance profile of an interacting quantum 
impurity}\label{app:cond_profile}
 As a reference to the results shown in Figs.\ \ref{fig:condd_vs_Eq} 
through \ref{fig:condq_vs_Ed}, in this section we show the typical 
zero--temperature behavior of the zero--bias conductance of a single 
interacting spin--1/2 quantum impurity, hybridizing with metallic leads. 
Parameters are $U_q = 0.5\,D$, $\Gamma_q = 0.05\,D$, with $D$ (the 
half--bandwidth of the leads' density of states) being used as an energy 
unit.
\begin{figure}%[t]
\begin{center}
\includegraphics[scale=0.33]{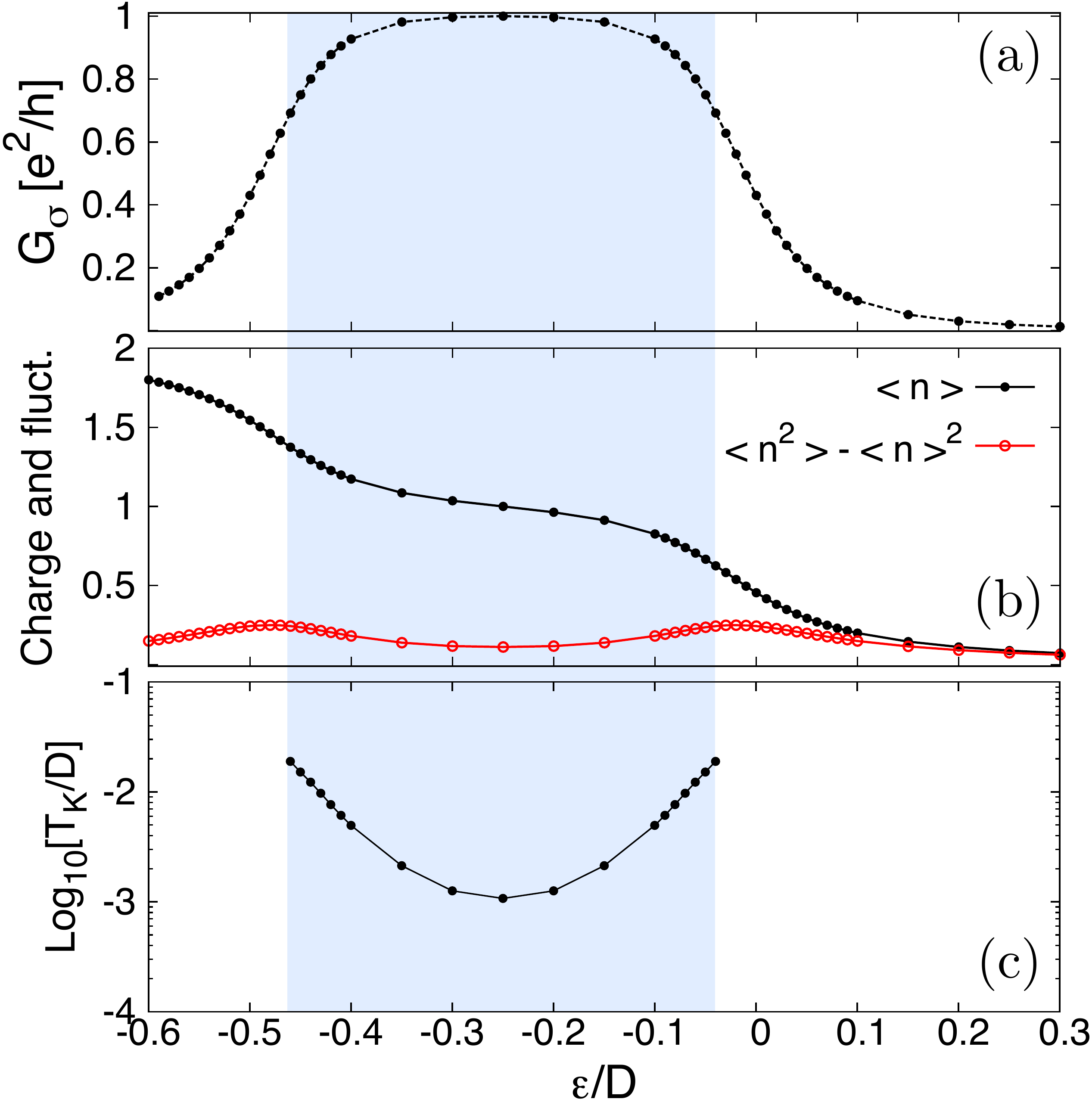}
\caption{(Color online) Typical (a) conductance (per spin channel) and (b) 
charge and charge fluctuation profiles (full and empty circles, 
respectively) of an interacting quantum impurity, as functions of the 
level 
energy $\varepsilon$. (c) Kondo temperature as a function of 
$\varepsilon$. 
The shaded region represents the range of $\varepsilon/D$ where the Kondo 
effect takes place. Curves calculated using NRG. Parameters: $U_q = 
0.5\,D$, $\Gamma_q = 0.05\,D$.}
\label{fig:charge_and_conductance_ref}
\end{center}
\end{figure}

Figure \ref{fig:charge_and_conductance_ref}(a) shows the variation of the 
conductance through the impurity: Coherent transport through the impurity 
is possible only when the impurity level is resonance with leads' Fermi 
level ($\varepsilon_F = 0$). The conductance is thus zero for 
$\varepsilon_q > 0$. Lowering the impurity level below zero, and within a 
range $\Gamma_q$ of the Fermi level, the conductance begins to rise as the 
system enters the mixed--valence regime. As the level goes lower and the 
average occupation of the impurity goes to one, the Kondo effect takes 
place, enhancing the conductance to the unitary limit ($G_{\sigma} 
\longrightarrow e^2/h$). The conductance then plateaus with the occupation 
$\xpect{n_{\sigma}}\approx 1$ [Figure 
\ref{fig:charge_and_conductance_ref}(b)] until $\varepsilon_q \approx -U_q 
+ \Gamma_q$, when the conductance falls again with increasing occupation 
of 
the impurity level, in accordance with Eq.\ (\ref{eq:friedel}).

\begin{acknowledgments}
The authors thank G. Martins for useful input at the 
beginning of this project. D. R. would like to thank A. Wong for his advice 
and for many useful discussions on the NRG approach to this problem. We 
acknowledge support by NSF MWN-CIAM grant DMR-1108285 and CONACyT 
(M\'exico). E.V. acknowledges support from CNPq, CAPES and FAPEMIG. S. E. 
U. thanks the hospitality of the Dahlem Center and the support of the 
Alexander von Humboldt Foundation. 
\end{acknowledgments}

\bibliography{bibliography.bib}{}
\end{document}